# Business Analysis: User Attitude Evaluation and Prediction Based on Hotel User Reviews and Text Mining

*Ruochun Zhao[1,a], Yue Hao[2,b], Xuechen Li[3,c]*

[1]Faculty of Information Science and Technology, Universiti Kebangsaan Malaysia, Bangi, Malaysia.
[2]Information Systems, Carey Business School, Johns Hopkins University, Baltimore, USA
[3]Schack Institute of Real Estate, New York University, New York, USA
a. zrc2664308120@gmail.com
b. yhao19@alumni.jh.edu
c. xl3281@nyu.edu

**Abstract.** In the post-pandemic era, the hotel industry plays a crucial role in economic recovery, with consumer sentiment increasingly influencing market trends. This study utilizes advanced natural language processing (NLP) and the BERT model to analyze user reviews, extracting insights into customer satisfaction and guiding service improvements. By transforming reviews into feature vectors, the BERT model accurately classifies emotions, uncovering patterns of satisfaction and dissatisfaction. This approach provides valuable data for hotel management, helping them refine service offerings and improve customer experiences. From a financial perspective, understanding sentiment is vital for predicting market performance, as shifts in consumer sentiment often correlate with stock prices and overall industry performance. Additionally, the study addresses data imbalance in sentiment analysis, employing techniques like oversampling and undersampling to enhance model robustness. The results offer actionable insights not only for the hotel industry but also for financial analysts, aiding in market forecasts and investment decisions. This research highlights the potential of sentiment analysis to drive business growth, improve financial outcomes, and enhance competitive advantage in the dynamic tourism and hospitality sectors, thereby contributing to the broader economic landscape.

# 1 Introduction

The rapid development of internet technology has profoundly transformed daily life, especially in sectors like e-commerce, which has significantly impacted industries such as tourism and hospitality. The hotel industry, which was severely disrupted by the global COVID-19 pandemic, has become a vital part of economic recovery in the post-pandemic period. The growing importance of user reviews has emerged as a key factor for assessing hotel services, especially as consumers increasingly rely on online feedback to evaluate the quality of hotels. Understanding consumer sentiment through review data has become essential for hotels to improve service and adapt to the changing market demand.

In the post-pandemic era, it is crucial for hotels to leverage advanced sentiment analysis techniques to better understand consumer emotions and expectations. By utilizing state-of-the-art natural language processing (NLP) technologies like BERT (Bidirectional Encoder Representations from Transformers) and deep learning, this study aims to extract actionable insights from hotel reviews, offering valuable guidance for service improvements and strategic decision-making. This approach not only helps in refining hotel service offerings but also contributes to broader market analysis, such as understanding consumer behavior and predicting financial trends.

A significant innovation of this research is the application of the BERT model, which offers superior accuracy in sentiment analysis compared to traditional methods. BERT's ability to understand contextual information and process data bidirectionally enables it to capture complex emotions expressed in user reviews, allowing for a more nuanced interpretation of customer sentiment. Unlike traditional approaches, which often focus on isolated words or phrases, BERT considers the entire context, improving the accuracy and depth of sentiment analysis. This improvement in sentiment classification enhances the model's adaptability to diverse and complex user feedback, making it more effective for real-world applications in the hospitality industry.

Additionally, this study addresses the common issue of data imbalance, especially in sentiment analysis where reviews tend to be polarized, often leaning either positive or negative. The research explores methods such as adjusting sample weights and employing oversampling or undersampling strategies to overcome this challenge. By implementing these techniques, the sentiment analysis model remains robust and accurate, ensuring its applicability in diverse situations where data may not be evenly distributed.

As the hotel industry quickly recovers in the aftermath of the pandemic, understanding user preferences and feedback has become crucial for staying competitive in the market. This research provides actionable insights that can help hotels enhance user satisfaction, differentiate themselves in a crowded marketplace, and improve service offerings. Furthermore, the insights gained from sentiment analysis are valuable for financial and market analysis, aiding investment strategies and the identification of growth opportunities. By integrating advanced NLP and sentiment analysis techniques like BERT, this study contributes to the development of cutting-edge methodologies in the field, offering a new approach to market research and consumer analysis in the hotel industry and beyond.

# 2 Literature review

## 2.1 Sentiment Analysis

Text sentiment analysis refers to the process of analyzing the sentiment expressed in textual data, identifying whether the sentiment is positive, negative, or neutral. In recent years, this technique has been widely applied to user reviews, providing businesses with valuable insights into customer attitudes. Given the diversity and evolving nature of human language, sentiment analysis techniques have evolved accordingly. Early methods relied on sentiment lexicons and syntactic structures, but modern approaches incorporate machine learning algorithms like Support Vector Machines (SVM), Naive Bayes, and more recently, deep learning models, which enable more nuanced sentiment classification. These advancements allow businesses to better understand customer expectations, improve service quality, and ultimately enhance customer satisfaction—especially crucial in industries like hospitality, where sentiment analysis can boost hotel occupancy rates and improve guest experiences.

Numerous scholars have contributed to the development of sentiment analysis methods. Yuan Xun, Liu Rong, and Liu Ming [1] proposed a BERT-based aspect-level emotion classification model that integrates multi-layer attention to improve sentiment classification by enhancing the attention on specific aspects and capturing long-term dependencies between aspect words and context.

Zhu Jian (2014) introduced methods based on sentiment lexicons, machine learning, and deep learning, highlighting the advantages and distinctions of each approach [2]. Zhou Yongmei, Yang Jianen, and Yang Aimin (2013) utilized HowNet and SentiWordNet models to generate more accurate sentiment lexicons for Chinese text [3]. Du Changshun and Huang Lei (2017) incorporated pooling and dropout algorithms to improve generalization ability and increase classification accuracy [4]. Chen Junqing and Zhang Yu (2018) combined the LSTM-LM model with neural networks to enhance sentiment analysis capabilities by capturing longer contextual information [5].

In the financial and economic context, sentiment analysis can also play a significant role in predicting market trends, as shifts in consumer sentiment often correlate with financial performance. This makes sentiment analysis a valuable tool not only for customer satisfaction but also for financial market forecasting and economic analyses. Techniques like PMI-IR dictionary expansion and statistical methods for rule extraction, as proposed by Ning X [6]. This growing body of research underscores the importance of sentiment analysis in both business and financial sectors, offering new ways to understand customer behavior and make data-driven decisions that improve market outcomes.

## 2.2 The current research status of the usefulness of user comments and online comments

In the era of the social Internet, online reviews have become not only a cultural phenomenon but also a significant form of word-of-mouth marketing, profoundly influencing consumer decision-making processes. This has led to increased attention from both academic researchers and industry practitioners, especially in the context of financial markets and economic analysis. Online reviews provide a rich source of alternative data that allows businesses to gauge consumer sentiment, which, in turn, can be leveraged for improving marketing strategies, investment decisions, and risk assessment.

Reviews have become a critical tool for shaping consumer perceptions, especially when it comes to making informed purchasing decisions. The influence of both positive and negative reviews is particularly pronounced in sectors such as e-commerce and logistics, where companies are increasingly leveraging sentiment analysis to refine their offerings and optimize customer experiences. Consumers are often inclined to scrutinize both positive and negative reviews to make more informed decisions, highlighting the importance of review sentiment in consumer behavior analysis. This trend has significant implications for financial technology, where businesses and investors alike use sentiment data to predict market trends, assess financial stability, and identify emerging investment opportunities.

Chen Ruixia and Li Yuanyuan suggest that factors such as brand reputation and product type are pivotal in determining the usefulness of reviews, influencing follow-up comments and overall review sentiment [7]. In a similar vein, studies by Wang X identify social benefits as a key driver of positive online reviews, particularly in the context of consumer engagement with online platforms [8]. The impact of review features such as image quantity and text length on customer satisfaction has also been widely studied, with research by Guo Yiqi et al [9]. revealing that these elements have a positive correlation with customer satisfaction, while excessive text length can negatively affect consumer perceptions.

From an economic perspective, reviews significantly influence market dynamics. For instance, Qian Lichun and Li Heng note that the quality, form, and value of reviews directly affect customer perceptions of risk and perceived usefulness, ultimately influencing their purchasing decisions [10].This is especially important in the context of financial markets where consumer sentiment, captured through review analysis, can serve as a leading indicator of broader market trends.

Furthermore, Zhao K., Stylianou A. C., and Zheng Y. argue that user reviews contribute to social influence and information flow in financial markets, where the perception of high-quality reviews can directly impact stock prices and investment decisions [11]. Jia Y. and Liu I. L. found that consumers prioritize the value of reviews—whether positive or negative—over the perceived utility, which is highly relevant for the evaluation of consumer behavior in financial technology markets [12].

# 3 BERT-based Sentiment Analysis Model

## 3.1 Introduction to Transformer Model

The Transformer is a neural network architecture used for natural language processing tasks. It was proposed by Google in 2017 and has achieved state-of-the-art results on various natural language processing tasks. The Transformer employs a sequence-to-sequence learning approach based on the attention mechanism, which calculates the relevance between words at different positions in the sequence and passes information between the encoder and decoder. The Transformer uses a multi-head attention mechanism, which splits the input into multiple subspaces and performs attention computation on each subspace to better capture semantic information. Additionally, the Transformer utilizes residual connections and layer normalization techniques to improve the convergence speed and stability of the model. It also introduces positional encoding to represent the positional information in the input sequence.

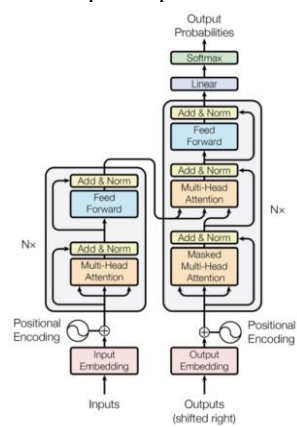

**Fig. 1.** The structure of Transformer [13].

The Transformer Encoder is a component of the Transformer model that is used to transform the input sequence into a series of hidden representations. It consists of multiple Encoder Layers, with each Encoder Layer containing two sub-layers: the multi-head self-attention layer and the fully connected feed-forward layer.

### 3.1.1 the multi-head self-attention mechanism layer

In the multi-head self-attention mechanism layer, each position in the input sequence is used to calculate its correlation with other positions in order to encode the dependencies between positions. This can be achieved by splitting the input into multiple sub-vectors, with each sub-vector used to calculate the self-attention mechanism. The layer takes the input sequence as queries, keys, and values.

Input into an attention mechanism, then multiply the attention weights with values and sum them to generate an output representation. The formula is as follows:

$$\text{MultiHead}(Q, K, V) = \text{Concat}(h_1, \ldots, h_h)W^O \quad (1)$$

$$h_i = \text{Attention}(QW_i^Q, KW_i^K, VW_i^V) \quad (2)$$

$$\text{Attention}(Q, K, V) \& = \text{softmax}\left(\frac{QK^T}{\sqrt{d_k}}\right)V \quad (3)$$

Among them, represents the number of heads, and is the weight matrix of linear transformation, which is the weight matrix obtained by concatenating multiple head results and using linear transformation to obtain the final output. It is the dimension of the key, which is used to scale the attention distribution in order to better control the range of the distribution.

*3.1.2 the multi-head self-attention mechanism layer*

In each Encoder Layer, the output of the multi-head self-attention mechanism layer is added and normalized through residual connections and layer normalization. Residual connection is the addition of input and output to reduce information loss, while layer normalization is the normalization of all feature dimensions of each sample to improve the stability and convergence speed of the model. The formula is as follows:

$$\text{LayerNorm}(x + \text{MultiHead}(Q, K, V)) \quad (4)$$

Where, x is the input vector.

*3.1.3 Feed Forward Networks*

In a fully connected feedforward layer, the output representation is transformed through a combination of two linear transformations and an activation function (usually). The purpose of this layer is to increase the non-linear and representational capabilities of the model, in order to better capture semantic information in the sequence. The formula is as follows:

$$\text{FFN}(x) = \max(0, xW_1 + b_1)W_2 + b_2 \quad (5)$$

Where $W_1, b_1$ and $b_2$ are the weight matrices and bias vectors of linear transformations.

Residual connections and layer normalization are also used in the fully connected feedforward layer. Specifically, it adds the output of the previous layer to the result after two linear transformations and activation functions, and then normalizes it through layer normalization. The formula is as follows

$$\text{LayerNorm}(x + \text{FFN}(x)) \quad (6)$$

A Transformer Encoder is a structure composed of multiple Encoder Layers that can transform input sequences into a series of hidden representations. Each Encoder Layer consists of a multi-head self-attention mechanism layer and a fully connected feedforward layer, and residual connections and layer normalization are used to improve the stability and representation ability of the model.

*3.1.4 Transformer decoder*

The decoder block of Transformer is also composed of six stacked decoders. Each decoder consists of a Masked multi-head attention mechanism, a multi-head attention mechanism, and a fully connected neural network. Compared to the encoder, it has an additional shielded multi-head attention mechanism, while the other structures are the same. The Masked multi-head attention mechanism masks certain values to prevent them from affecting parameter updates. The output of the converter is first linearly transformed, and then the probability distribution of the output is obtained through the softmax function. Finally, the word with the highest probability in the dictionary is selected as the predicted output.

### 3.2 Word vectorization based on BERT model

BERT is a Transformer based pre-trained language model that obtains a universal language representation through two stages of training. In the pre-training stage, a large amount of unlabeled text data is used for training, and context semantics and sentence associations are learned through Masked Language Model (MLM) and Next Sentence Prediction (NSP) tasks. In MLM tasks, the model predicts masked words through context, thereby learning to understand missing words in the context. In NSP tasks, the model predicts the coherence and semantic association between two sentences, improving the model's ability to understand sentence-level semantics.

The input of BERT includes the initial word vector, text vector, and position vector. The initial word vector is randomly initialized, the text vector is used to capture the global semantic information of the text, and the position vector is used to represent the semantic information differences carried by words at different positions. Through pre-training and fine-tuning, BERT can be applied to various natural language processing tasks, such as text classification, named entity recognition, and question-answering systems. The performance of the BERT model has made significant breakthroughs in the field of natural language processing and has become one of the most advanced natural language processing models at present.

Compared with traditional Attention mechanisms, the BERT model considers the order and position information of each word. This approach solves the problem of ignoring the order of words or characters in traditional Attention mechanisms.

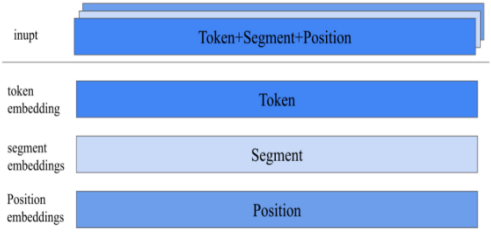

**Fig. 2.** BERT Model Input [14].

Through the two tasks of MLM and NSP, the BERT model can pre-train on a large amount of text data, learn deep representations of language, and thus perform well in various natural language processing tasks.

In the BERT model, input data undergoes three important steps of processing to obtain an effective text vector representation. Firstly, in the Token Embedding stage, each word or tag is transformed into a fixed dimensional embedding vector to capture the semantic relationships between words. Next is Segment Embedding, which is used to distinguish the semantics and associations between different sentences. Finally, there is Position Embedding, which considers the order of words in the text. After completing these three steps, the text vector representation is obtained. Then, this text vector is further processed through multiple Transformer encoder layers. Each encoder layer includes a Self-attention mechanism for modeling text associations and enhancing semantic expression. After undergoing Self-attention, semantic vectors are processed through a series of operations and activation functions to extract rich semantic features [10]. Then, an enhanced semantic vector is obtained through linear transformation and random deactivation operation, which is used as input for the next encoder. This process is repeated in the remaining encoder layers, giving the BERT model strong semantic expression capabilities. This multi-layered encoder structure makes BERT a highly effective and widely used model in natural language processing tasks.

## 4 Results

### 4.1 Evaluation of Model Related Indicators

Calculate the accuracy, precision, recall, F1 score, and other metrics of the model on the test set to determine its performance. The commonly used indicators for evaluating the effectiveness of model training are accuracy and log loss. The higher the accuracy, the better the classification performance of the model.

(1) Precision rate (check rate): The precision rate represents the proportion of positive cases among all samples predicted to be positive cases. In other words, it measures how accurately the model predicts positive cases.

$$Precision = \frac{TP}{TP+FP} \quad (7)$$

(2) Recall rate: Recall rate represents the proportion of samples that are actually positive and correctly predicted by the model as positive. It reflects the coverage of the model for positive samples.

$$Recall = \frac{TP}{TP+FN} \quad (8)$$

(3) F1 value: F1 value is a comprehensive indicator of accuracy and recall, used to balance the relationship between the two. The higher the F1 value, the better the model performs in terms of accuracy and recall.

$$F1 = \frac{2TP}{2TP+FP+FN} \quad (9)$$

(4) Accuracy: Accuracy refers to the proportion of correctly classified samples in all predicted results to the total number of samples. However, in cases of imbalanced samples, accuracy is not a good evaluation indicator.

$$Accuracy = \frac{TP+FN}{TP+FP+TN+FN} \quad (10)$$

The adjustment of model parameters is achieved through backpropagation of the loss function value, which measures the difference between the predicted and actual values of the model. Therefore, the smaller the logarithmic loss value, the better the classification performance of the model. By setting the number of training iterations (epoch=5), the model can be trained multiple times, gradually optimizing the model. The following figure shows the variation curves of accuracy and recall during the training process.

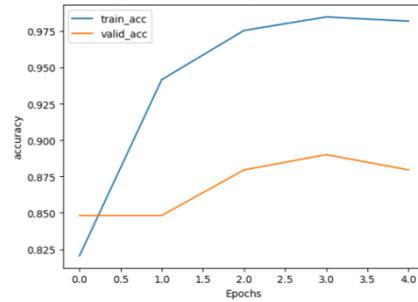

**Fig. 3.** Changes in classification accuracy of the BERT model.

From the above figure, it can be seen that as the number of model training increases, when the epoch reaches 4, the accuracy of the BERT model gradually increases, approaching 1, and finally reaches 0.8795. The logarithmic loss value decreases from 0.4555 to 0.3403, approaching 0, indicating that the model has been effectively trained. The following is the confusion matrix for different label pairs.

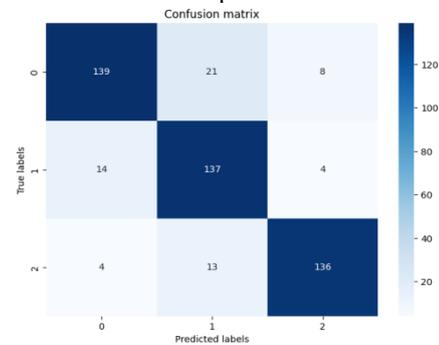

**Fig. 4.** BERT model confusion matrix diagram.

In deep learning, the balance of data distribution is generally adjusted during the experimental process based on empirical judgment. In the task of deep learning comment classification, different data distributions can lead to different prediction results. The main parameter to adjust is the balance of data distribution.

During the model training process, the model parameters are adjusted through backpropagation to minimize the difference between the predicted values and the actual values, i.e., the loss function value. Therefore, a smaller logarithmic loss value corresponds to better classification performance. To achieve better performance, multiple training iterations are needed. By setting the number of training iterations (epoch) to 10, the model can achieve better performance after multiple iterations.

## 5 Conclusions

This study conducted an in-depth analysis of hotel review data, building an effective rating classification model based on sentiment analysis, and fine-tuned the BERT model to achieve significant classification results. The main findings of the study include: successfully overcoming the issue of data imbalance, especially when the imbalance ratio was 30% and 40%, where the model performed optimally with high precision, recall, and F1-score; excellent classification results were achieved on the test set through training and evaluation of the BERT model; and the model, when applied to real business data, provided valuable insights regarding customer satisfaction, key issues, competitor analysis, and market trends, offering a decision support tool for hotel managers. However, there are some limitations in the study, mainly: limited data size, which may not cover all scenarios and types of reviews. Future research could expand the data set to verify the robustness and generalization capability of the model; sentiment labeling could be more detailed, as the current study used only positive, neutral, and negative labels, and further refinement could provide more specific sentiment information; additionally, BERT and other deep learning models have challenges with interpretability, and future work should aim to improve model interpretability to better meet practical business needs.

This study offers significant insights for the business sector, especially in the hotel industry and other service industries. The sentiment analysis model helps businesses gain a comprehensive understanding of user sentiments towards their products and services. By identifying and addressing negative emotions in a timely manner, user experience can be enhanced. Furthermore, sentiment analysis can guide the optimization of service quality, including improving facilities, enhancing hygiene management, and optimizing customer service. Through competitor review data analysis, businesses can grasp the market competition landscape and adjust their strategies. In addition, large-scale review data analysis helps detect market trends and changes in consumer preferences, enabling businesses to adjust products, services, and marketing strategies. In the future, sentiment analysis will have broader and deeper applications in the business domain. Research can explore more advanced natural language processing models, such as GPT series, and combine multimodal data for joint analysis to improve sentiment analysis accuracy. Furthermore, the model can be extended to other industries, such as dining, retail, and tourism, with customized development. Real-time sentiment analysis systems will enable businesses to quickly respond to market changes, enhancing their competitive advantage. Sentiment analysis visualization tools will also help non-technical personnel understand the analysis results and promote better decision-making.